\documentstyle[aps,prl]{revtex}
\begin{document}
\twocolumn[
\title{Calculation of eigenvalues of a strongly chaotic system using 
Gaussian wavepacket dynamics}

\author{Arjendu K. Pattanayak$^{1,2}$ and William C. Schieve$^1$} 

\address{1)Ilya Prigogine Center for Studies in Statistical Mechanics \& 
Complex Systems,\\ Department of Physics, The University of Texas, Austin, 
TX 78712 \\ and \\ 2\cite{tor}) Chemical Physics Theory Group, University of 
Toronto,\\ Toronto, Ont. M5S 3H6, Canada }
\date{\today}

\maketitle
\widetext 
\begin{abstract}
{
We apply the approximate dynamics derived from the Gaussian time-dependent 
variational principle to the Hamiltonian
$ \hat H= \frac{1}{2}(\hat p_x ^2+ \hat p_y ^2)+ \frac{1}{2}\hat x^2\hat y^2$, 
which is strongly chaotic in the classical limit. We are able to calculate, 
essentially analytically, low-lying eigenvalues for this system.
These approximate eigenvalues agree within a few percent with the numerical 
results. We believe that this is the first example of the use of TDVP dynamics 
to compute individual eigenvalues in a non-trivial system and one of the few 
such computations in a chaotic system by {\em any} method. 
}
\end{abstract}
\pacs{PACS numbers: 05.45.,03.65.S,05.40}
]
\narrowtext

\section{Introduction}
Gaussian approximations to quantum mechanics have been utilized successfully
in many contexts including quantum field theory\cite{everyone},
the dynamics of hydrogen plasma\cite{klakow}, semiclassical propagation
methods\cite{heller,pre}, quantum control\cite{wilson} and the 
study of ``quantum chaos"\cite{leshouches}. 
The primary motivation for their popularity is simplicity of computation:
Gaussians are easily parameterized by the $c$ number variables 
specifying the centroid (average variables) and spread (fluctuation 
variables) and their dynamics are essentially classical, apart from the 
computation of a phase which is a crucial element. Further, 
Gaussians arise naturally in the coherent state representation of quantum 
mechanics\cite{barg} and in the $N=\infty$ limit (where $N$ is number of 
degrees of freedom) of many-body systems\cite{yaffe}. There are a variety
of Gaussian approximations including: (1) a variational approximation 
usually derived through the time dependent variational principle 
(TDVP)\cite{everyone,Dirac,KraSa}, (2) a recently introduced quadratic-order 
Gaussian approximation\cite{pre}, (3) Heller's method\cite{heller}, which is 
a non-self-consistent truncation of (2), and 
(4) the multiple classical trajectory version of Heller's 
method\cite{tomsovic}. A further level of approximation yields the 
Gaussian Effective Potential method, which consists of an 
adiabatic elimination of the time-dependence of the fluctuation variables 
in the TDVP dynamics\cite{stev}.

The wide-spread use of these methods raises the question of their validity 
and range of applicability\cite{hage}. 
It has been argued that in the presence of chaos, semiclassical approximations 
to quantum mechanics should break down on a logarithmic time-scale 
$t_c \sim \frac{1}{\lambda}\log(\frac{1}{\hbar})$, where $\lambda$ is the 
largest Lyapunov exponent of the underlying classical mechanics, and $\hbar$ 
is Planck's constant. Computations with the multiple trajectory Gaussian 
approximation has demonstrated that this may be a pessimistic 
estimate\cite{tomsovic}. Recently, the validity of the TDVP approximation has 
also been considered. Apart from the chaos in the underlying 
classical system, it has been shown that the approximate quantum dynamics 
derived from the TDVP may be chaotic even when the classical limit is 
not\cite{prl}. This has led to the argument\cite{bala} that the TDVP 
Gaussian approximation fails in the presence of chaos. However, it has been 
shown that this anomalous chaotic behavior persists even when exact numerical 
computations are made\cite{larry1}. It is argued, in fact, that this chaos is 
a signature of the complicated nature of the spectrum involved in the exact 
quantum dynamics\cite{pre,larry2}.
Further, recent work by Habib\cite{salman} shows that {\em all} Gaussian 
approximations to Schrodinger's equation are identical to the same
approximation to the {\em classical} Liouville equation, although the 
classical versions do not have any phase information.
This result clarifies that $\hbar$ is a kinematical constant in 
these approximations, providing a scale for the ``smoothing" of the dynamics 
and reinterprets the ``quantum effects" included in Gaussian approximations. 
It does not invalidate the results of the quantum 
Gaussian approximations, although it does require these results to be 
understood in kinematical rather than dynamical terms. 
The result also emphasizes that quantum dynamics are better approximated by 
classical Liouville dynamics rather than Hamilton's equations for point 
trajectories\cite{liov}. 

It is thus clear that Gaussian approximations should be used and interpreted 
with caution. However, as we shall demonstrate in this paper, the TDVP 
Gaussian approximation {\em does} yield accurate results even in the presence 
of chaos, in a system where other approximations fail. The TDVP dynamics can 
be used to compute eigenvalues\cite{Kan,pre} through an extension of 
the Einstein-Brillouin-Keller quantization method\cite{EBK}. We use this 
method to compute eigenvalues for the two-dimensional coupled quartic 
oscillator
\begin{equation}
\hat H = \frac{\hat p_x ^2+ \hat p_y ^2}{2} + \frac{1}{2}\hat x^2\hat y^2.
\label{eq:ham}
\end{equation}
This system is highly chaotic classically\cite{saviddy}; it was believed till 
recently to be ergodic, and the integrable regions of phase space occupy less 
than $0.005\%$ of the volume. 
The quantal Hamiltonian also resists numerical analysis; large basis-sets do 
not suffice for quantization\cite{Eckhardt}. It is usual for numerical ease to 
add a term like 
$\beta(\hat x^4 + \hat y^4)$ to the potential and to analyze the system in 
the limit $\beta \to 0$. The traditional methods for semiclassical 
quantization fail for this particular system. Firstly, the EBK method cannot 
be applied because of the chaos in the system -- there are no stable torii
to use for quantization. The alternative methods developed in the field
of quantum chaos\cite{leshouches} also do not work. The most prominent of 
these, the Trace Formula method, starts from Feynman's path-integral 
representation of quantum mechanics, and through a sequence of stationary 
phase approximations, derives the eigenvalues of the quantum system as poles 
in a weighted sum over the unstable periodic orbits of the classical 
system\cite{gutz}. 
Recent successes in other systems\cite{gutz,wintgen} notwithstanding, the
systematic enumeration of the unstable periodic orbits of the classical 
system has not been achieved for this particular system. Another alternative, 
Heller's approach\cite{heller}, which computes an approximate time-dependent 
wave function $\langle\Psi(t)\vert$ and Fourier transforms the overlap 
$\langle\Psi(t)\vert\Psi(0)\rangle$, fails because of the inherent instability 
of the truncation, as argued in a previous qualitative\cite{pre} analysis in 
model potentials. In the same paper, we 
demonstrated that the TDVP dynamics restores stability to systems where the
truncated Gaussian dynamics fail; in fact, the TDVP dynamics can be stable
even when the {\em classical} dynamics are unstable. This is precisely what 
happens in the system given by Eq.~(\ref{eq:ham}): An infinite set of 
unstable periodic orbits is stabilized by the ``quantal fluctuation" terms in 
the TDVP method. These orbits can then be used to compute approximate 
eigenvalues for a symmetry subspace of this system. Remarkably, these are 
obtained {\em analytically } (barring a numerical integral). 
Further, these results are obtained with $\beta =0$, 
i.e. the regime where even large basis-set calculations fail to converge. 
The high degree of chaos in the system and the fact that we use $\hbar=1$ in 
our calculations suggests the n\"aive perspective that the classical-like 
TDVP approach is then far from its region of validity. However, our results 
are extremely accurate when compared with the ``exact" numerical results over 
a substantial range for the lowest-lying eigenvalues. 
Thus, while there is no suggestion that this method can be always used in the 
presence of chaos to successfully approximate quantum dynamics, our results 
indicate that it can certainly be used with care in some circumstances.

In the next section, we shall briefly review the dynamical equations for the 
TDVP method, including the construction of a quantization rule. 
In the third section, we apply this method to Eq.~(\ref{eq:ham}), comparing 
the method in the process to the usual semiclassical methods. We then discuss 
the results and argue that the regime of validity of the method is the 
low quantum-number regime, contrary to the usual intuition derived from the 
Correspondence Principle that classical-like approximations work best in the 
high quantum number regime\cite{leshouches}. We thus suggest that the TDVP 
Gaussian approximation works best as a technique complementary to the usual 
semiclassical methods.

\section{TDVP Gaussian dynamics}

The usual derivation of these dynamics proceeds from Dirac's Time-Dependent
Variational Principle (TDVP)\cite{Dirac,KraSa}; this posits an action of the 
form $\Gamma =\int dt\langle\Psi,t\vert i\hbar{\partial\over\partial t}-
\hat H|\Psi,t\rangle$.
The general requirement that $\delta \Gamma = 0$ yields the Schr\"odinger
equation and its complex conjugate. The true solution is approximated by
restricting $\vert\Psi(t)\rangle$ to a subspace of the full Hilbert space and 
setting $\delta \Gamma = 0$ within this subspace. In particular, this 
restriction may be to the space of general Gaussians\cite{everyone}. 
We have derived these same dynamics\cite{prl,pre} in a somewhat more 
intuitive fashion from Ehrenfest's Theorem. We start with the equations for 
the centroid variables and make a Taylor expansion around the centroid 
with the higher moments of the wave function. These moments follow the usual 
Heisenberg dynamics, yielding in general an infinite system of equations.
We render this system finite by projecting onto the space of Gaussians;
this system of equations are identical to those derived from the TDVP.

We have been able to represent these dynamics as an extended {\em classical
gradient system} for the average and fluctuation variables with dynamical
equations 
\begin{eqnarray}\label{eq:ext_dyn1}
{d x \over dt} &= & p ,\\
\label{eq:ext_dyn2}{d p\over dt}  &= &
 - \sum_{m=0}^{m=\infty} {\rho ^{2m} \over m! 2^m} V^{(2m+1)} (x), \\
\label{eq:ext_dyn3}{d \rho \over dt}  &= & \Pi, \\
{d \Pi\over dt} &= &{ \hbar^2  \over 4 \rho ^3}
-\sum_{m=1}^{m=\infty}{\rho^{2m-1} \over (m-1)!2^{m-1}}V^{(2m)}(x), 
\label{eq:ext_dyn4}
\end{eqnarray}
and a Hamiltonian
\begin{eqnarray}
& H_{ext} & = {p^2 \over 2} + { \Pi ^2 \over 2 } + V_{ext}(x,\rho);\\
V_{ext} (x,\rho) = & V(x) &  + {\hbar ^2\over 8 \rho^2} +
\sum_{m=1}{\rho^{2m} \over m!2^m }V^{(2m)}(x)
\end{eqnarray}
where the subscript {\em ext} indicates the ``extended" potential and 
Hamiltonian and $V^{(n)} = {\partial ^n V / \partial x^n} |_{<\hat x>}$.
The coordinate variables for the extended Hamiltonians are $x,\rho$ and their 
canonically conjugate momenta are $p,\Pi$ and are related to the
moments of the Gaussian $\Psi(x,p,\rho,\Pi,t)$ as follows:
\begin{eqnarray}
\langle\hat x\rangle &\equiv& x \label{eq:x},\\
\langle\hat p\rangle &\equiv& p, \\
\langle\Delta x \Delta p + \Delta p \Delta x\rangle &\equiv& 2\rho \Pi,
\label{eq:xp}\\
\langle\Delta x^{2m}\rangle &=&{(2m)!\rho^{2m}\over m!2^m}\label{eq:gau1},\\
\langle\Delta x^{2m+1}\rangle &=& 0 \label{eq:gau2},\\
\rho^2\langle\Delta p^2\rangle &=&  \frac{\hbar^2}{4} + \rho^2 \Pi^2.
\label{eq:unc}
\end{eqnarray}
The first 3 relationships Eqs.~(\ref{eq:x} -- \ref{eq:xp}) are 
{\em definitions} and Eqs.~(\ref{eq:gau1} -- \ref{eq:gau2}) are a consequence
of the Gaussian {\em ansatz}.  Eq.~(\ref{eq:unc}) is a kinematical constraint 
arising entirely from choosing the equality in the Uncertainty Principle 
relationship; it is the only way in which $\hbar$ enters this approximation.

In this method, the fluctuation and average variables are treated on the same 
footing and the phase space is dimensionally consistent: $\rho$ 
has the dimensions of length and $\Pi$ that of momentum. The geometry of the 
space is thus identical to that of an ordinary classical system -- it is a 
Cartesian space, a manifold ${\cal R}(2N)$ defining the extended 
phase-space. There is also an equation for the phase of the wave function: 
If we define $\lambda$ as
$\vert\Phi, t\rangle\equiv\exp\left(\frac{i\lambda (t)}{\hbar}\right)
\vert\Psi,t\rangle$
it is simple to derive from Schr\"odinger's equation the equations
for $\lambda = \lambda _D + \lambda _G$ with the first part
\begin{equation}
\lambda _D = - \int_0^t d\tau\;\langle\hat H\rangle = -t \;H_{ext}
\end{equation}
corresponding to the dynamical phase. The second part is the geometrical phase:
\begin{equation}
\lambda _G  = \int_0^t d \tau\; \langle i\hbar {\partial \over \partial 
\tau}\rangle 
  =  \int_0^t d \tau \;({\dot \rho \Pi - \dot \Pi \rho \over 2}+ p\dot x ).
\label{eq:lg1}
\end{equation}
For cyclic evolution this is the Aharanov-Anandan form of 
``Berry's phase"\cite{berry}; it depends only on the geometry of the 
evolutionary path in phase space and can be written as
\begin{equation} \label{eq:lg}
\lambda _G (C) = \oint _C {\bf P \cdot d Q},
\end{equation}
where ${\bf P}\equiv (p,\Pi)$ and ${\bf Q}\equiv (x,\rho)$.
The equation for the phase, along with the Hamiltonian equations of motion for
the evolution of the wave function parameters constitute the TDVP dynamics.
This lies on a space ${\cal R}(2N) \times S (1)$; for the
case just considered, $N = 2$, the result is completely general, however.

We now provide a constructive argument\cite{pre} for obtaining eigenfunctions 
and eigenvalues, which is equivalent to imposing a single-valuedness 
constraint on stationary wavefunctions\cite{Kan} in the extended
phase-space. Note that an eigenfunction for the extended dynamics is one
whose parameters are {\em invariant} under the evolution.
We see readily that a periodic orbit (PO) solution to Hamilton's equations
Eqs.~(\ref{eq:ext_dyn1}-\ref{eq:ext_dyn4}) is invariant on the ${\cal R}(2N)$ 
subspace; however, each point along the PO acquires a phase factor during 
the evolution. The dynamical phase is the same for all the points along the 
PO and can be factored as a global phase. The geometrical part $\lambda _G$ 
for the PO is crucial: 
We note that a $PO \times \lambda _G$ such that the periodic evolution
of $\lambda _G$ on ${\cal S}(1)$ is {\em commensurate} with that of the PO on
${\cal R}(2N)$ is a function invariant on entire space ${\cal R}(2N) \times
{\cal S}(1)$ and is hence an {\em eigenfunction}. The commensurability of 
the phase translates to the relation
\begin{equation}
\lambda _G (PO) =\frac{1}{2\pi} \oint _{PO} {\bf P \cdot dQ} = n \hbar ,
\end{equation}
where we have used Eq.~(\ref{eq:lg}). 
Thus, the eigenfunction is a weighted sum (the weight factor at each point 
being the appropriate geometrical phase) over the points of the commensurate 
periodic orbit and the eigenvalue is $H_{ext}$ for that PO. 
This rule is the same as the ``old" quantization rule of Bohr and
Sommerfeld; however, it applies in the {\em extended} phase space, as opposed 
to the classical phase space and thus does not have the same meaning. 
In particular, there are no Maslov-Morse corrections\cite{EBK} to this rule, 
since there are no singularities in the Gaussian representation. It can be 
shown\cite{lj,pre} that the ``spread" variables $\rho,\Pi$ explicitly take 
care of these corrections. In general, this quantization condition will give 
results different from the EBK rule (the POs are in the extended space) but 
always incorporates the Maslov correction.
The extension of this argument from POs to invariant torii goes through
easily\cite{Kan} and leads to a general quantization rule:
\begin{equation}\label{eq:pdq}
\frac{1}{2 \pi}\oint _{C_i} {\bf P \cdot dQ} =  n_i\hbar
\end{equation}
where the closed integral is now taken over the $i$th irreducible contour 
around the torus and the quantum numbers $n_i$ are labeled accordingly. 
This is exactly Einstein's generalization \cite{Einstein} of the 
Bohr-Sommerfeld rule to invariant torii. 

The system of equations derived by Heller\cite{heller} for the semiclassical 
evolution of Gaussian wavepackets obtain as truncations of equations 
(\ref{eq:ext_dyn1}) and (\ref{eq:ext_dyn2}) to ${\cal O}(\rho ^0)$ and of 
equations (\ref{eq:ext_dyn3}) and (\ref{eq:ext_dyn4}) to ${\cal O}(\rho ^1)$.
This arguably\cite{pre} inconsistent semiclassical system of 
equations destroys the Hamiltonian structure of the dynamics, leading to 
non-unitary evolution\cite{Lj}. A consistent truncation to ${\cal O}(\rho ^1)$) 
for this system retains the Hamiltonian structure of the TDVP method and has 
been termed extended semiclassical dynamics\cite{pre}: All the advantages
of the TDVP method applies to the extended semiclassical method, including the 
definition of a Poisson bracket, and the existence of a unitary propagator and 
an analytic quantization method. Unlike the TDVP method, 
Heller's Gaussian dynamics and the extended semiclassical dynamics arise as 
``controlled" first-order expansions; their validity can thus be formally 
evaluated\cite{hage} and these are hence attractive approximations.
However, the truncation of the dynamical equations at the term involving
the third derivative of the potential induces an unphysical instability, 
where the fluctuation variables grow without bound\cite{pre} even for simple 
one-dimensional anharmonic potentials like $V(x) = x^4$. 
In the TDVP dynamics, all orders of derivatives are maintained with a 
re-summation of the moment expansion under a Gaussian {\em ansatz}; this 
yields behavior that is qualitatively similar to the exact long-term quantal 
behavior, in particular reproducing the appropriate stability. Thus, the 
TDVP method can stabilize unstable periodic orbits, which may then be used 
as above to obtain eigenvalues and eigenfunctions, as we now demonstrate.

\section{Eigenvalues for a chaotic Hamiltonian}
We now turn to the computation of eigenvalues for the Hamiltonian
\begin{equation}
\hat H = {\hat p_x^2 + \hat p_y^2 \over 2} + {1 \over 2}\hat x^2 \hat y^2
+ \beta(\hat x^4 +\hat y^4).
\end{equation}
Extensive numerical work\cite{Eckhardt,saviddy} shows that the classical 
limit ($\hat O \to O$ for all operators) is a very strongly chaotic system, 
with few stable periodic orbits in the limit $\beta \to 0$.
It is easy to verify, however, the existence of an infinity of unstable 
periodic orbits along the diagonals of the potential. We also note here that 
the Hamiltonian displays a simple scaling relationship in 
energy\cite{Eckhardt}: 
A trajectory $({\bf x}_1(t),{\bf p}_1(t))$ at an energy $E_1$ is
related to a trajectory $({\bf x}_2(t),{\bf p}_2(t))$ at an energy $E_2$ by
\begin{eqnarray}
{\bf x}_2(\tau) &=& (\frac{E_2}{E_1})^{\frac{1}{4}}{\bf x}_1(t)\\
{\bf p}_2(\tau) &=& (\frac{E_2}{E_1})^{\frac{1}{2}}{\bf p}_1(t)
\end{eqnarray}
where $\tau$ is the rescaled time $\tau = (\frac{E_2}{E_1})^{-\frac{1}{4}}.$
This means that there is the same degree of strong chaos at all finite 
energies: There is no ``transition to chaos". 
Eckhardt, Hose and Pollack have done a careful numerical analysis of the 
quantum system to show the presence of ``scars" in the 
eigenfunctions\cite{Eckhardt}. They state that harmonic oscillator basis-set 
quantization with matrices of dimension $3240$ do not provide converging 
eigenvalues for $\beta=0$; they have hence used $\beta =0.01$ for their 
analysis. The eigenfunctions of this Hamiltonian belong to the symmetry 
classes of the $C_{4\nu}$ symmetry group which has 
eight elements (four reflections in the axes and diagonals and four rotations 
by $\pi/2$). The irreducible representations of this group split into four 
one-dimensional representations and one two-dimensional representation.
They have restricted themselves to the four one-dimensional representations 
corresponding to wave functions which are 
A) symmetric under $x \to y, x \to -x$,
B) antisymmetric, symmetric, C) symmetric, antisymmetric and
D) antisymmetric, antisymmetric respectively, and have
numerically obtained low-lying eigenvalues and eigenstates for this system.

We have applied the Gaussian wavepacket methods detailed above to this 
system with $\beta=0$. Of the three methods, the truncated Gaussian methods 
(both Heller's dynamics and the extended semiclassical system) fail, 
yielding unstable motion where the wavepackets spread without bound, 
irrespective of the value of $\beta$. This is easily established by noting 
that there exist one-dimensional projections in which this two-dimensional 
potential reduces to the anharmonic quartic potential considered above 
and the existence of a single unstable direction for the dynamics corresponds 
to instability in the global motion. 
This exposes one particular frailty of the truncated Gaussian approximations: 
they work well in systems that are close to harmonic only in the 
{\em particular} sense of being potentials of the form $V(x) = x^2 + f(x)$ 
with the function $f(x)$ containing higher polynomials.

On the other hand, the TDVP method works excellently for this system; firstly, 
the dynamics are completely bounded. It is an interesting feature of this 
approach that even though there exist classically unbounded orbits along 
$x=0$ or $y=0$ (which are precisely what make the numerical quantal analysis 
through basis sets so difficult) the inclusion of the re-summed 
moment terms via the variational approach restores stability to the problem. 
This effect has been termed ``quantum resuscitation" in the context of the 
Gaussian Effective Potential\cite{stev}.
The extended Hamiltonian for the TDVP (using $\Psi = \Phi(x)\Phi(y)$) is
\begin{eqnarray}
H_{ext} & =&  {1 \over 2}(p_x^2 +p_y^2 + \Pi_x^2 + \Pi_y^2)\nonumber\\
 & + &   {1 \over 8 \rho _x^2} + {1 \over 8 \rho _y^2}
+ { 1 \over 2}(x^2 + \rho _x^2)(y^2 + \rho _y^2).
\end{eqnarray}
Note that the use of the factored wave functions explicitly restricts us
to the one-dimensional representations.
The dynamics of this extended Hamiltonian are, in general, chaotic. However,
if we consider the subspaces of the one-dimensional representation noted
above, we find that the first and third subspace can be studied by the 
symmetry-reduced version of $H_{ext}$:
\begin{equation}
H = { 1\over 2}(p^2 + \Pi ^2) + {1 \over 8 \rho ^2} +
{1 \over 4}(z^2 + \rho ^2)^2
\end{equation}
where $(z,p)$ and $(\rho, \Pi)$ are the canonically conjugate pairs.

We now demonstrate that this symmetry-reduced version of the {\em extended}
Hamiltonian is explicitly integrable. To do so, we make the change of 
variables to spherical coordinates $R,\theta$ defined in the plane 
by: $z = R \cos \theta ; \rho = R \sin\theta$ .
This transforms the Hamiltonian to
\begin{equation}
H = \frac{p_R^2}{2} + \frac{R^4}{4} +  \frac{1}{R^2}
\left [ \frac{p_{\theta}^2 }{2} + \frac{1}{8 \sin^2\theta} \right].
\end{equation}
Since the factor multiplying $\frac{1}{R^2}$ is solely a function of $\theta$,
this is now in the right form to use Hamilton-Jacobi theory \cite{LandL}.
Through the separability just noted, therefore, we introduce Hamilton's
characteristic
functions $W_R$ and $W_{\theta}$ and get the Hamiltonian-Jacobi equations:
\begin{eqnarray}
\frac{1}{2} \left (\frac{\partial W_{\theta}}
{\partial \theta}\right)^2 + \frac{1}{8 \sin^2\theta}&=& k \\
\frac{1}{2} \left ( \frac{\partial W_R}{\partial R}\right)^2 + \frac{R^4}{4}
+ \frac{k}{R^2} &=& E
\end{eqnarray}
where $E$ (the energy) and $k$ are the constants of separation.
We form action variables as usual:
\begin{eqnarray}
J_{\theta} &=& \frac{1}{2\pi}\oint d\theta \frac{\partial W_{\theta}}
{\partial \theta}\\
&=&\frac{1}{2\pi}\oint d\theta \sqrt{2k - \frac{1}{4 \sin^2\theta}}\\
J_R &=& \frac{1}{2\pi}\oint dR \frac{\partial W_R}{\partial R} \\
    &=& \frac{1}{2\pi}\oint dR \sqrt{2E - \frac{2k}{R^2} -\frac{R^4}{2}}.
\end{eqnarray}
The $\theta$ integral yields\cite{rhyzik} 
\begin{equation}
J_{\theta} = \frac {\alpha -1}{2}
\end{equation}
where $\alpha$ is introduced for convenience through 
$k\equiv\frac{\alpha ^2}{8}.$ The $R$ integral is a complicated elliptic 
integral that can be evaluated; however, it cannot be analytically inverted 
to yield the quantization condition. We hence leave it in quadrature:
\begin{equation}
J_R =\frac{1}{2\pi}\oint dR\sqrt{2E-\frac{(2J_{\theta}+1)^2}{4R^2}-
\frac{R^4}{2}}.
\end{equation}
The existence of this integral demonstrates the integrability of the chosen
symmetry subspace of the extended Hamiltonian. Since we now have a set of
stable invariant torii in the extended space, we can proceed with the
quantization as detailed above in a straightforward fashion. To wit: 
Eigenvalues correspond to torii with quantized actions in both variables $R$ 
and $\theta$. We follow this prescription by setting $J_{\theta}$ equal to a 
half-integer in the above equation [the symmetry-reduced form of the 
Hamiltonian only accumulates half the phase of the actual Hamiltonian which 
is why we use half-integers];
this yields a one-degree of freedom dynamical system in $R$ which has only
closed orbits. We then proceed as follows: We take various initial 
conditions and numerically integrate their dynamics over the closed orbit to 
compute the action $\oint P_R dR$. The orbits for which the action equals 
a half-integer then correspond to eigenfunctions and their conserved energy 
the associated eigenvalue. We show the results for the first $67$ eigenvalues 
in Fig. \ref{fg:Eck}, compared with the numerical results\cite{rescale} of 
Eckhardt {\em et al}. We now note that the eigenvalues we have calculated 
essentially analytically agree within a few percent with the numerical 
results\cite{Eckhardt} over the substantial range of our calculations.

We emphasize that there is no possibility that the validity of the results 
can be attributed to the minute regions of stability of the classical
phase-space\cite{Dahl}. 
A moment's consideration shows that the use of the factored wave functions 
and the restriction $y(t)=x(t)\equiv z(t)$ for the symmetric dynamics 
corresponds to restricting our attention to the diagonals of the potential. 
That is, the dynamics of the Gaussian are restricted such that the centroid 
always travels always along the unstable periodic orbits along these 
diagonals. The few classical stable periodic orbits that do exist in this 
system lie far from this region and our results cannot be understood as 
affected by the presence of these orbits; the wavepacket is not influenced 
by them. 

Further, we note that there is a superficial similarity of these results to 
other work\cite{bender} that demonstrates the efficacy of Gaussian 
approximations in computing low-lying spectral features. However, those
results depend on perturbations of a classically integrable Hamiltonians.  
As such, they were able to use standard quantal perturbation theory. 
This is not possible for Eq.~(\ref{eq:ham}). Further, since the classical 
dynamics is {\em ab initio} strongly chaotic, even the Gustavson-Birkoff 
quantization technique\cite{delos}, which is an adaptation of classical 
perturbation theory, cannot be applied. In either case, there is nothing to 
perturb around for this system. This also emphasizes the non-perturbative 
aspect of the variational approximation.

It is clear that the TDVP method in this system takes advantage of the 
interesting feature of being able to use the infinity of periodic orbits 
along the diagonals of the potential. These orbits are classically 
{\em unstable}, and the formal application of the WKB quantization method 
to this unstable periodic orbit yields metastable states\cite{miller}, where 
the eigenvalues have a real part (corresponding to the energy) and an 
imaginary part (corresponding to the life-time of the state). Apart from
the unphysical metastability of these states, the approximate energies thus 
obtained are valid only for the first few states -- we show the limited 
accuracy in Fig.~1, where we have plotted the real parts of the first $10$ 
eigenvalues from this method. This same instability of the periodic orbit 
causes the breakdown of the truncated Gaussian semiclassical methods as well. 
However, the TDVP Gaussian {\em ansatz} makes dynamics along these periodic 
orbits {\em stable} -- an example of ``quantum resuscitation" -- and our 
generalized quantization method succeeds. 

There are more general considerations also: 
Note that, despite the high degree of chaos in the system, the 
potential is relatively benign for Gaussian approximations: It has only one 
minimum and no maxima. However, since it is a quartic well, this does not
benefit the truncated approximations and we must turn to the fully re-summed 
TDVP approximation to take advantage of this structure. Second, the high 
value of $\hbar$ is an advantage in this context. 
As recent work\cite{korsch} demonstrates, at higher values of 
$\hbar$ the details of the classical phase-space structure are smoothed out
in the quantal dynamics (see figure (2a) of \cite{korsch}), as reflected by 
the quantal eigenfunctions. Since $\hbar$ also sets the kinematical scale of 
smoothness for the TDVP Gaussian method\cite{salman}, this means that both 
the exact quantal dynamics and our approximate version are effectively 
occurring in a smoother potential well than the classical point dynamics. 
Both these factors imply that distorted Gaussians can be expected to 
evolve without excessive error under such circumstances. Further, the 
smoothness permits the weighted superposition of Gaussians inherent in the 
TDVP eigenfunction {\em ansatz} to yield accurate results: 
The detailed structure of the wave functions may be argued to contribute 
rapidly oscillating terms that affect the computation of {\em averages} of 
observables in these states in a small way. Thus, eigenvalues, which are the 
average of the Hamiltonian operator, may be quite accurate even when the 
detailed dynamics of the variational approximation are not so. This 
also clarifies that the method is expected to be valid in the regime of the 
lowest-lying eigenvalues, contrary to the usual Correspondence Principle 
regime where classical-like approximations work best in the high quantum 
number regime\cite{leshouches}. This is supported by our results: they 
deviate slowly away from the numerical results as the quantum number 
increases.

Our results thus show that the degree of chaos in a classical problem does
{\em not} necessarily limit the validity of the TDVP Gaussian dynamics.
In a general case, chaos in the classical dynamics of a system coexists with 
complicated potential energy surfaces; Gaussian methods would hence apply 
only over a small range of parameters. However, there are systems that may be 
extremely chaotic, but possess the appropriate structure that enables the 
TDVP Gaussian approximation to work over a much larger range, especially in 
the high $\hbar$ regime; this explains the many successful 
applications\cite{klakow}, for instance. It is clear that these arguments 
need to be explored carefully in more situations. 

In summary, we have analytically (barring a numerical integral) computed
the lowest-lying eigenvalues of a classically strongly chaotic system. 
We believe that this is the first example of the use of TDVP dynamics to 
compute individual eigenvalues in a non-trivial system and one of the few such 
computations in a chaotic system by {\em any} method. These 
results compare extremely favorably with numerical results and show that the 
limits of validity of the TDVP method are not necessarily set by the degree 
of chaos in the classical system. Accurate quantum dynamical simulations are 
quite difficult and there is a great need for valid approximations\cite{QMD}. 
Heller's truncated Gaussian approximation has already been shown\cite{heller} 
to be extremely useful for systems that can be explicitly written as
perturbations around a harmonic minimum. It is intuitive that Gaussian
approximations should continue have validity in potential wells, even in 
the presence of chaos; however, the truncated Gaussian methods fail in 
anharmonic systems. We believe that the TDVP Gaussian method is an excellent
candidate for approximate quantum dynamics in these and other situations.

\acknowledgements
It is a pleasure to acknowledge many fruitful interactions with Bala Sundaram 
and Salman Habib; we thank Dr. Habib for a detailed discussion of unpublished 
results. The Robert A. Welch Foundation (Grant F-0365) and the Natural 
Sciences and Engineering Research Council of Canada provided partial support 
for A.~K.~P. during this work.

\begin{figure}[htbp]
\caption{Comparison of lowest eigenvalues}\label{fg:Eck}
\end{figure}


\begin{thebibliography}{99}
\bibitem{rescale}We note here that the numbers reported in \cite{Eckhardt} 
are scaled eigenvalues. A failure to note this lead to a mistaken 
statement in\cite{pre} that this method improved upon the numerical results 
of\cite{Eckhardt}; as demonstrated in this paper, our results agree very well 
with the numerical values.
\end{thebibliography}
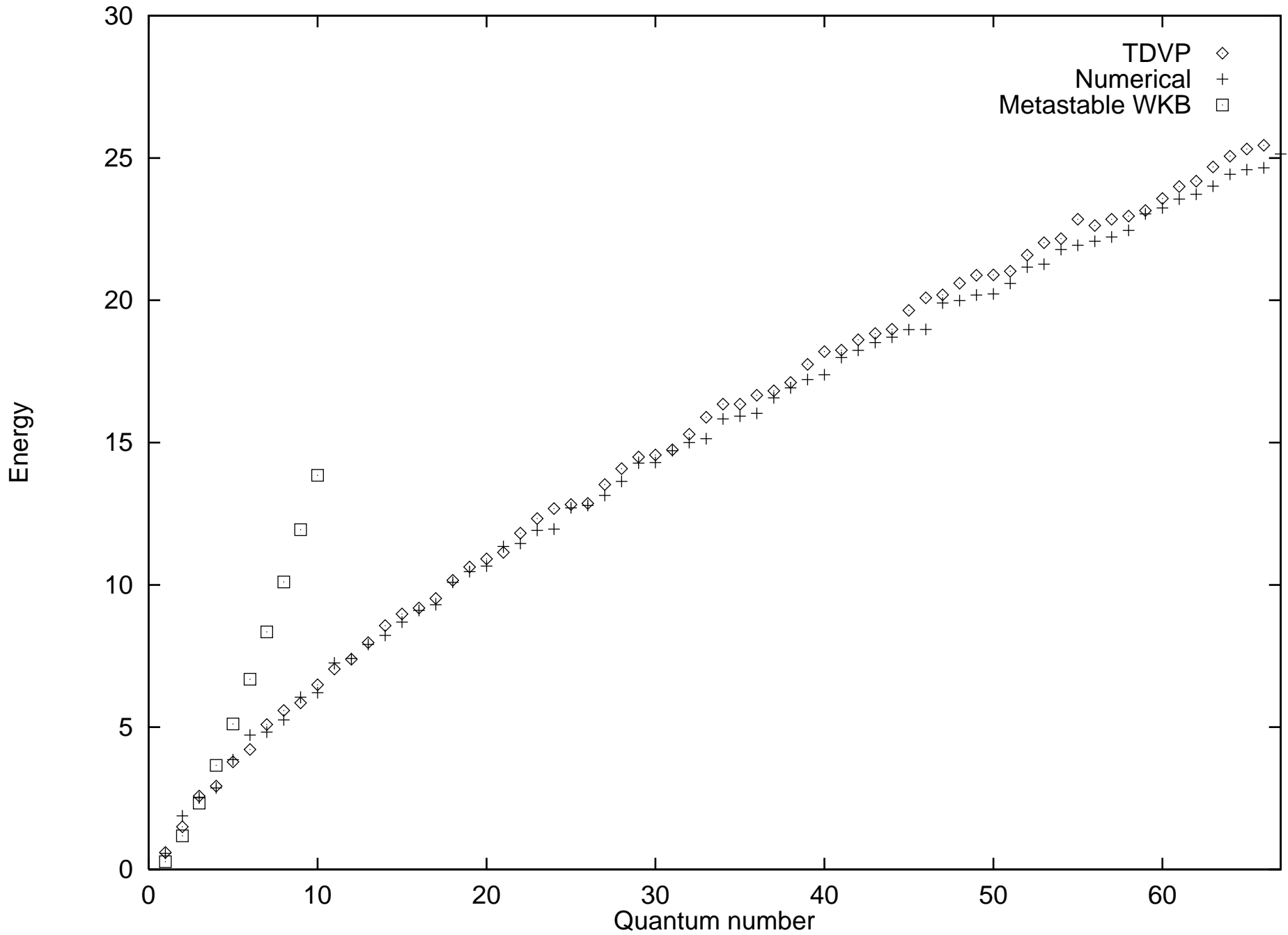
\end{document}